\documentclass[runningheads]{llncs}
\usepackage[T1]{fontenc}
\usepackage{graphicx}
\usepackage[hidelinks]{hyperref}
\usepackage{color}

\usepackage{algorithmic}
\usepackage{textcomp}
\usepackage{xcolor}
\usepackage{xspace}
\usepackage[per-mode=symbol,mode=text,detect-all]{siunitx}

\usepackage{fontawesome}
\usepackage{amssymb}
\usepackage{rotating}
\usepackage{tabularx}
\usepackage{booktabs}
\usepackage{adjustbox}

\usepackage[inline]{enumitem}
\usepackage{cleveref}
\usepackage[T1]{fontenc}

\usepackage[misc]{ifsym} % \Letter command

\renewcommand{\orcidID}[1]{\href{https://orcid.org/#1}{\includegraphics[width=8pt]{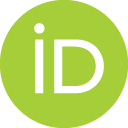}}}

\newif\ifdraft
\newif\ifanonymous
\newif\ifauthor

\drafttrue
\draftfalse
\anonymoustrue
\anonymousfalse

\authortrue
\newcommand{\ie}{i.e.,\xspace}
\newcommand{\eg}{e.g.,\xspace}

\makeatletter
\newcommand{\printfnsymbol}[1]{%
  \textsuperscript{\@fnsymbol{#1}}%
}
\makeatother

\ifauthor
	\usepackage{tikz}
	\newcommand\copyrighttext{
	\footnotesize
	\hspace{5pt}

	This version of the contribution has been accepted for publication, after peer review but is not the Version of Record and does not reflect post-acceptance improvements, or any corrections.
	Use of this Accepted Version is subject to the publisher's Accepted Manuscript terms of use \url{https://www.springernature.com/gp/open-research/policies/accepted-manuscript-terms}.
	}

	\newcommand\copyrightnotice{%
		\begin{tikzpicture}[remember picture,overlay] \node[anchor=south,yshift=40pt] at (current page.south)
		{{\parbox{\dimexpr\textwidth-\fboxsep-\fboxrule\relax}{\copyrighttext}}};
		\end{tikzpicture}%
	}
\fi

\begin{document}
\title{Reputation Systems for Supply Chains:\\The Challenge of Achieving Privacy Preservation}
\titlerunning{Reputation Systems for Supply Chains}
\ifanonymous
	\author{Anonymous Author(s)}
	\authorrunning{Anon. Author(s)}
\else
	\author{%
	\href{mailto:lennart.bader@fkie.fraunhofer.de}{\Letter}~Lennart Bader\inst{1}\orcidID{0000-0001-8549-1344}\thanks{The authors contributed equally to this work.} \and
	\href{mailto:pennekamp@comsys.rwth-aachen.de}{\Letter}~Jan Pennekamp\inst{2}\orcidID{0000-0003-0398-6904}\printfnsymbol{1} \and
	Emildeon Thevaraj\inst{2}\orcidID{0009-0005-9225-5397} \and\\
	Maria Spiß\inst{3}\orcidID{0000-0002-1645-4635} \and
	Salil S.\ Kanhere\inst{4}\orcidID{0000-0002-1835-3475} \and
	Klaus Wehrle\inst{2}\orcidID{0000-0001-7252-4186}
	}
	\authorrunning{L.\ Bader et al.}
	\institute{%
	Cyber Analysis \& Defense, Fraunhofer FKIE, Germany
	\and
	Communication and Distributed Systems, RWTH Aachen University, Germany
	\and
	Institute for Industrial Management at RWTH Aachen University, Germany
	\and
	School of Computer Science and Eng., University of New South Wales, Australia
	}
\fi

\maketitle              % typeset the header of the contribution

\begin{abstract}
Consumers frequently interact with reputation systems to rate products, services, and deliveries.
While past research extensively studied different conceptual approaches to realize such systems securely and privacy-preservingly, these concepts are not yet in use in business-to-business environments.
In this paper,
\begin{enumerate*}[label=(\arabic*)]
	\item we thus outline which specific challenges privacy-cautious stakeholders in volatile supply chain networks introduce,
	\item give an overview of the diverse landscape of privacy-preserving reputation systems and their properties, and
	\item based on well-established concepts from supply chain information systems and cryptography, we further propose an initial concept that accounts for the aforementioned challenges by utilizing fully homomorphic encryption.
\end{enumerate*}
For future work, we identify the need of evaluating whether novel systems address the supply chain-specific privacy and confidentiality needs.
%
%\keywords{supply chain management \and confidentiality \and voter \and votee anonymity \and fully homomorphic encryption}
\end{abstract}
\section{Introduction and Motivation}
\label{sec:introduction}

\ifauthor
	\copyrightnotice
\fi
In contrast to historically long-living business relationships, today's supply chain networks are much more volatile~\cite{Pennekampetal2019Dataflow}.
This shift manifests in both
\begin{enumerate*}[label=(\roman*)]
	\item the demand for more flexible, spontaneous and short-lived business agreements and
	\item a growing demand for digitized relationship establishments and management.
\end{enumerate*}
While these aspects are well-known influencing factors for traditional supply chain management (SCM)~\cite{Pennekampetal2024An}, where various tracing systems tailored to the specific privacy and transparency needs of businesses within and along supply chain structures have been proposed~\cite{Baderetal2021Blockchain-Based,Gonczoletal2020Blockchain}, these solutions are not applicable for easing the business partner selection processes in such volatile supply chains.

For spontaneously establishing new relationships, potential business partners are in need of reliably assessing the credibility, \ie{} \emph{reputation}, of each other.
Similarly, they have an incentive to advertise their own business and services to potential partners, where a well-built reputation might attract further business and sales.
Since these use cases require businesses to transparently provide insights into their operations and potentially their former business relationships, confidentiality concerns on business and production secrets must be considered at all times when discussing and proposing respective technical solutions~\cite{Pennekampetal2024An}.

The concept of measuring a business' or product's reputation and granting access to it to potential customers or partners is well-established~\cite{Hendrikxetal2015Reputation}, \eg{} for online sellers and their offered articles.
However, since such reputations are mostly based on non-verifiable reviews, they can neither guarantee reliability nor accuracy, limiting their value for customers and the businesses themselves.

Hence, business-to-business (B2B)-focused reputation systems, as prevalent in the context of supply chains, exhibit a distinct set of requirements regarding
\begin{enumerate*}[label=(\arabic*)]
	\item the reliability of ratings and the resulting reputation,
	\item the transparency properties,
	\item the privacy of involved stakeholders, and
	\item the confidentiality of business information.
\end{enumerate*}
Even though related work studies reputation systems~\cite{Gurtleretal2021SoK:,Hasanetal2022Privacy-Preserving}, they largely fail to consider the specific needs of B2B settings.
In this paper, we thus raise the awareness for this research gap by identifying and discussing respective requirements along with potential conflicts.
Moreover, we propose a first concept for a privacy-preserving multi-agent B2B reputation system that combines well-established concepts from supply chain information systems and cryptography for a flexible trade-off between these requirements.

\section{Reputation Systems}
\label{sec:background}

Information \emph{transparency} is an important aspect of today's supply chain networks~\cite{Pennekampetal2024An}.
Businesses have identified the various benefits of inter-business transparency for enhancing their collaboration for increased success of both their individual business and their supply chain as a whole~\cite{Baderetal2021Blockchain-Based}.
However, \emph{privacy} preservation is an equally important aspect as businesses want to keep their business relationships a secret.
Consequently, generally maintaining the \emph{confidentiality} of information (except for deliberately shared slices) as well as sufficient \emph{security} mechanisms (\eg{} to protect against unauthorized access as well as manipulation) are fundamental requirements for supply chain-oriented information systems.

\subsection{Related Work: The Diversity in Today's Reputation Systems}
\label{subsec:background:relatedwork}

The universal benefits of reputation systems have resulted in diverse approaches for various domains.
Due to our focus on privacy preservation and confidentiality in supply chains, we study relevant approaches from two well-known surveys~\cite{Gurtleretal2021SoK:,Hasanetal2022Privacy-Preserving} in light of these requirements, \ie{} we omit inapplicable approaches from our analysis.
We augment this foundation with a selection of recent papers~\cite{Maliketal2019TrustChain:,Zhouetal2021Blockchain-based,Putraetal2022DeTRM:,Arshadetal2022REPUTABLEA} to provide an up-to-date overview.
In \Cref{tab:rw} (Appendix), we detail the specific features of these approaches, which serve as the basis for our own design.
Just like previous conclusions~\cite{Gurtleretal2021SoK:}, we confirm that today's systems are unable to satisfy all desirable combinations of properties.
Especially privacy properties often depend on the inclusion of at least one trusted third party, which is a challenging and potentially unrealistic assumption for supply chain settings.

\subsection{Requirements for Privacy-preserving Reputation Systems}
\label{subsec:background:researcharea}

The desire for inter-business transparency, aiming at an accountable selection of new business partners and increasing their own business' visibility, leads to the research area of privacy-preserving reputation systems~\cite{Gurtleretal2021SoK:,Hasanetal2022Privacy-Preserving}.
These reputation systems adduce customer and business feedback in the form of \emph{votes} to derive a business' \emph{reputation} as a measure of its trustworthiness, quality of offered services, and ability to cooperate.
Other customers and businesses can then request this reputation as a foundation for their follow-up decision-making.

Hence, businesses can take the role of a \emph{voter} when submitting a vote or rating, a \emph{votee} when receiving such ratings, or a \emph{requester} when requesting access to the derived reputation score, as formalized by Gurtler and Goldberg~\cite{Gurtleretal2021SoK:}.
Each business taking one or multiple roles introduces specific requirements to the reputation system.
Straightforward requirements, \ie{} the reliability and accountability of the reputation, the security and flexibility of submitted ratings, and the possibility for access control for reputation requests, are complemented by the demand for the complex trade-off between privacy and transparency~\cite{Baderetal2021Blockchain-Based}, especially since ``privacy'' can have different notions depending on a business' role~\cite{Gurtleretal2021SoK:}.
Common privacy requirements for voters are related to their anonymity and the anonymity of individual (\emph{voter-vote privacy}) and even multiple (\emph{two-vote privacy}) votes~\cite{Gurtleretal2021SoK:}.
Further, votees might require privacy (confidentiality) regarding their reputation, either by proving their current reputation without the need to link themselves to a long-term history (\emph{reputation-usage unlinkability}) or by providing meta-information on their current reputation, \eg{} threshold-based, instead of their plain reputation score (\emph{exact reputation blinding})~\cite{Gurtleretal2021SoK:}, raising reputation reliability challenges.
Similarly, requesters might want to hide their identity when requesting another business' reputation, which, among other aspects, might contradict the votees' demand for sophisticated access control.

The variety of contradicting requirements poses a significant challenge for researchers when designing appropriate and practical reputation systems.
Each assessment of the importance or trade-off weights of these requirements leads to different potential designs, such that a wide range of systems has been proposed.

\section{Reputation and Privacy Preservation in Supply Chains}
\label{sec:problemstatement}

The lack of widely-used reputation systems in supply chain settings hints at issues with today's technical approaches (cf.\ \Cref{tab:rw} for general academic concepts).
To better assess this inadequate situation, we now explore the circumstances of deploying and developing supply-chain-focused reputation systems.

\subsection{Supply Chain-induced Requirements}
\label{subsec:problemstatement:domainrequirements}

In contrast to commonly studied consumer-oriented reputation systems, \eg{} seller or product ratings on online platforms, reputation systems for B2B use and supply chains face specific confidentiality and privacy needs, which render the application of existing reputation systems difficult or simply infeasible.

First, supply chains induce new \emph{information flow dimensions}~\cite{Pennekampetal2024An}.
Combining subjective ratings with contract-based information and objective production- or service-related information for computing a reputation score requires suitable reputation functions, authenticity checks, and verification mechanisms.

Second, the volatility of modern supply chain networks is a big challenge because, \eg{} systems with \emph{votee-owned}~\cite{Gurtleretal2021SoK:} scores are not applicable when participants go out of business or just decide to stop participating.
Likewise, selecting a trusted third party for managing the businesses' reputations that all participants agree on represents a significant challenge for globalized supply chains.

\subsection{Research Gap: Domain-specific Realization}
\label{subsec:problemstatement:researchgap}

Existing (privacy-preserving) reputation systems (cf.\ \Cref{subsec:background:relatedwork}) fail to consider the previously outlined requirements regarding privacy, availability, and the supported data dimensions.
For a formal definition of these dimensions in supply chains, we refer to Pennekamp et al.~\cite{Pennekampetal2024An}.
Hence, developing a sophisticated privacy-preserving design for all involved parties that does not require well-established trust between participants while also accounting for volatile structures remains an open issue in research.
In the following, we outline promising design decisions for such a reputation system and detail its conceptual design.

\section{Toward a Comprehensive Design}
\label{sec:design}

Based on the general requirements for privacy-preserving reputation systems and the additional challenges arising from the specifics of modern supply chain networks, we now derive a comprehensive design for a reputation system tailored to these challenging needs.
Given our business focus, we know that all entities (operators, participants, and users) are bound by legislation and to specific jurisdictions.
Hence, for this work, we consider malicious-but-cautious attackers~\cite{Ryan2014Enhanced} who can misbehave in all possible ways while trying to not leave any verifiable evidence of their misbehavior.
With this attacker model, we have to include more attacks than with honest-but-curious (semi-honest) attackers since it explicitly allows for local deviation from protocols unless they are provable by third parties.
Consequently, collusion attacks are possible but attackers would have to make sure to not leave any (public) traces to comply with the attacker model.

\subsection{Design Decisions}
\label{subsec:design:buildingblocks}

Achieving a reasonable trade-off between reputation privacy and the desired degree of transparency while maintaining a secure operation and accounting for volatile supply chain structures calls for
\begin{enumerate*}[label=(\arabic*)]
	\item a sophisticated encryption scheme,
	\item verifiable reputation computations, and
	\item a distribution of competences to multiple independent entities (as we assume malicious-but-cautious adversaries).
\end{enumerate*}

First, we identify a \emph{ticket-based}~\cite{Gurtleretal2021SoK:} voting process as well suited for combining privacy preservation with rating reliability:
A business should be authorized to submit a vote for another business only if a (recent) business relationship existed.
Hence, voting tickets can be issued as soon as a business relationship has been established, \ie{} a new contract has been signed.
The tickets can either be issued by the businesses themselves or by a third entity that reviews relationships within an already-deployed supply chain information system, such as the privacy-preserving realizations ProductChain~\cite{Maliketal2018ProductChain:} or PrivAccIChain~\cite{Baderetal2021Blockchain-Based}.

Second, to account for the challenge of dealing with volatile supply chain structures (especially, defunct, \ie{} departing, businesses), we assess that a \emph{centralized} approach fits best to ensure a reliable operation of the system (availability of reputation scores).
With a (conceptual) central and static entity, departing or non-collaborative nodes do not negatively affect the reputation system's operation.
Using multiple entities for different roles and potentially splitting a single role onto multiple entities further reduces the risk of operational deficiencies.

Third, as we already encourage the use of \emph{multiple independent entities} for different roles, we reduce the trust in a single entity and limit its ability to compromise information privacy and security, \ie{} preventing the threat and impact of collusion attacks.
While such an approach already achieves full collusion resistance, increasing the number of central entities further distributes critical responsibilities, which allows for a tunable degree of collusion resistance.

Fourth, to achieve privacy-preserving voting and request processes, we propose the use of \emph{pseudonyms}.
Instead of authorizing themselves to other businesses or the reputation engine, voters and requesters receive a (temporary) pseudonym from an independent or government-run entity to authorize their requests.
Thus, individual requests cannot be linked to a specific business, and multiple requests from the same business appear as if they were coming from different entities, such that the businesses' privacy as well as privacy of relationship can be achieved, accounting for the needs of (volatile) supply chain networks.

Finally, the use of a (conceptually) central entity mandates an elaborate data security concept.
Specifically, we propose to base the voting process as well as the reputation function on \emph{fully homomorphic encryption} (FHE)~\cite{Marcollaetal2022Survey}.
In particular, voters encrypt their votes under FHE before submitting them to an available reputation engine, which is unable to decrypt or tamper with individual votes.
The reputation engine then computes the reputation score under FHE while offering a verification mechanism to both voters and votees.
Since the result remains encrypted, the reputation engine does not learn the actual, potentially sensitive reputation score, ensuring the desired votee privacy.

To quickly explore the feasibility of FHE for our use in reputation systems, we measured the runtime of the basic operations needed in our design using Pyfhel~\cite{Ibarrondo2017Pyfhel}, which is a Python library for Microsoft SEAL~\cite{Microsoft-Inc.2018Microsoft}, utilizing the CKKS~\cite{Cheonetal2017Homomorphic} scheme.
We extrapolated these numbers to realistic scenarios, and after discussions with our applied partner, we concluded that FHE is suitable.

\subsection{The Rating Process on a High-level}
\label{subsec:design:designoverview}

\begin{figure}[t]
  \center
  \includegraphics[width=0.8\linewidth]{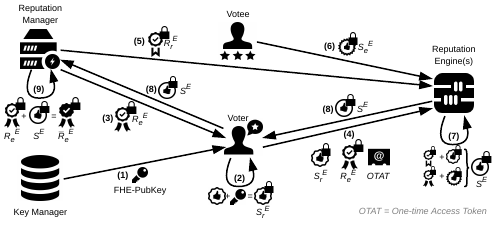}
  \vspace{-1.5em}
  \caption{The reputation calculation is shielded using FHE to ensure confidentiality.}
  \vspace{-1.5em}
  \label{fig:rating-process}
\end{figure}

To illustrate our design and the interaction of its entities, we now present the rating process consisting of multiple steps (cf.\ \Cref{fig:rating-process}).
For the rating, we combine and weigh subjective ratings with a rating based on objective information on the business relationship.

During the rating process, both the \emph{voter} and the \emph{votee} use (temporary) pseudonyms for their interactions.
The \emph{voter}
\begin{enumerate*}[label=(\arabic*)]
  \item retrieves the \emph{votee}-specific FHE key from the \emph{key manager} and
  \item encrypts its rating with this key under FHE as $S^E_r$. The voter then
  \item receives the (FH-encrypted) current reputation $R^E_e$ of the \emph{votee} from the \emph{reputation manager} to
  \item forward $S^E_r$, $R^E_e$, and a one-time \emph{access token} to one of the available \emph{reputation engines}. The \emph{reputation engine} then
  \item uses the access token to receive the \emph{voter's} current reputation $R^E_r$ from the \emph{reputation manager} and
  \item optionally receives a data-backed (which ideally is verifiable, \eg{} since it has been processed using trusted computing~\cite{Pennekampetal2020Secure}) self-rating $S^E_e$ from the \emph{votee} to
  \item combine $S^E_r$ and $S^E_e$ into the final rating $S^E$. Here, the current reputations $R^E_r$ and $R^E_e$ can be used as weights for the ratings. The \emph{reputation engine} then
  \item signs and submits the new rating to the \emph{voter}, who forwards the rating $S$ back to the \emph{reputation manager}. Here, the new reputation
  \item is calculated as $\bar{R}^{E}_e$ using $R^E_e$ and $S^E$.
\end{enumerate*}

The use of FHE ensures that neither the reputation engine nor the reputation manager can access individual ratings or a business' plain reputation.
By distributing the reputation computation to multiple entities, \ie{} \emph{reputation engines}, we maintain privacy of relationship between voter and votee toward the \emph{reputation manager}.
The use of the one-time \emph{access token} further prevents the \emph{reputation manager} from linking \emph{voter} and \emph{votee} by request analyses, as the current reputations are requested by different entities.

\section{Conclusion and the Road Ahead}
\label{sec:conclusion}

In this paper, we have highlighted the lack of reputation systems in B2B settings and volatile supply chain networks despite the availability of technical approaches, indicating a mismatch between offered features and set requirements.

\textbf{Evaluation Challenges.}
Generally, this research is hindered by the lack of evaluation data and realistic supply chain models~\cite{Pennekampetal2024An}.
While our applied partner (Institute for Industrial Management) can provide us with real-world data, its scale and the complexity of business relationships represent only the tip of the iceberg.
Specifically, at this point, we rely on data from ERP systems that cover the order and delivery of goods.
Moreover, we have access to monitoring data of production machines that allow us to incorporate such data into the calculation of reputation.
Regardless, we are interested in acquiring additional datasets to extensively evaluate reputation systems for volatile supply chain networks.

\textbf{Future Work.}
In addition to evaluating our proposed approach, we further identify three crucial research directions:
\begin{enumerate*}[label=(\arabic*)]
	\item Assessing whether multi-key FHE can offer confidentiality benefits,
	\item studying how to better utilize existing supply chain information systems, and
	\item investigating the implications of dealing with E2E-secured supply chains~\cite{Pennekampetal2020Secure}, \ie{} incorporating trusted sensors into the reputation computation and capitalizing on their reliability benefits.
\end{enumerate*}

\textbf{Conclusion.}
The lack of deployed reputation systems in B2B environments encouraged us to look into the respective reasons since reliable reputation scores are beneficial for businesses when managing both short- and long-term operations.
Even though privacy-preserving concepts are available, their use is hindered by volatile supply chain networks and differing data dimensions.
Addressing this research gap, we are proposing an FHE-based design that accounts for the transparency, confidentiality, and privacy requirements of participating businesses while being compatible with existing supply chain information systems.

\subsubsection*{Acknowledgments.}
Funded by the Deutsche Forschungsgemeinschaft (DFG, German Research Foundation) under Germany's Excellence Strategy -- EXC-2023 Internet of Production -- 390621612 and the Alexander von Humboldt (AvH) Foundation.

\bibliographystyle{splncs04-etal}
\bibliography{paper}

\appendix
\section{Appendix: Comparing Reputation Systems in Detail}

As a foundation of our work (cf.\ \Cref{subsec:background:relatedwork}), we have analyzed various reputation systems, especially those covered in previous surveys~\cite{Gurtleretal2021SoK:,Hasanetal2022Privacy-Preserving}, regarding their system architecture and their properties for both feedback provision and the reputation itself.
In \Cref{tab:rw}, we summarize the corresponding results.

For the architecture, we distinguish between \textit{centralized}~(C), \textit{decentralized}~(D), and \textit{hybrid}~(H) approaches.
Additionally, depending on the system, voters can provide feedback values from different sets or value ranges.
Here, we indicate the set of available values either as a set (\eg{} \(\mathbb{R}\) or \(\{1, ..., 5\}\)) or as an interval.
Some systems also allow text and vectors as feedback.
Moreover, the feedback granularity indicates whether feedback is provided for a \textit{single}~(S) interaction between voter and votee or over \textit{multiple}~(M) ones.

For reputation, we further cover six distinct properties per approach.
While the set or range is equivalent to the respective feedback property, liveliness~\cite{Schiffneretal2009Privacy} indicates whether negative ratings are allowed by the system.
This aspect is also related to monotonicity, which indicates that a votee's reputation can only increase over time.
Hence, non-monotonicity allows for reputations to decrease over time or when negative ratings are submitted.
Some reputation systems provide \textit{global}~(G) visibility, while others offer \textit{local}~(L) visibility.
With global visibility, all requesting parties receive the same response to a reputation query.
With local visibility, different requesters can receive different results.
The durability indicates whether ratings are stored locally or whether the reputation has to be recalculated on every request, resulting in a trade-off between storage and computation requirements.
Finally, a reputation system's aggregation model indicates how feedback is aggregated into a final reputation score.

%!TEX root = ../paper.tex

\newcommand{\ayc}[3]{#1 (#2)~\cite{#3}}
\newcommand{\mymidrule}{\cmidrule(lr){1-10}}

\def\arraystretch{1.5}%  1 is the default, change whatever you need
\newcolumntype{Y}{>{\centering\arraybackslash}X}
\newcolumntype{R}[2]{%
    >{\adjustbox{angle=#1,lap=\width-(#2),margin=0em 0em 0em 0em}\bgroup}%
    c%
    <{\egroup}%
}
\newcommand*\rot{\multicolumn{1}{R{45}{1em}}}%
\newcommand*\rotb[1]{\rot{\textbf{#1}}}%
\renewcommand\tabularxcolumn[1]{m{#1}}

\begin{table}[ht]
\tiny
\centering
\caption{%
We analyzed privacy-preserving reputation systems that are promising for use in supply chains regarding nine distinct properties.
We represent no level of fulfillment of the respective property by \faCircleO{}\;while \faCircle{} shows fulfillment of the property.
For entries labeled with ``?'', we cannot reliably identify the corresponding aspect.%
}
%\vspace{-0.85em}

\setlength{\tabcolsep}{0pt}
\begin{tabularx}{\textwidth}{
@{\hspace{1.1em}}>{\hangindent=1em\raggedright\hsize=3\hsize}Y
>{\hsize=.5\hsize}Y
>{\hsize=1\hsize}Y
>{\hsize=.5\hsize}Y
>{\hsize=1\hsize}Y
>{\hsize=.5\hsize}Y
>{\hsize=.5\hsize}Y
>{\hsize=.5\hsize}Y
>{\hsize=.5\hsize}Y
>{\hsize=2.9\hsize}Y
}

\toprule
\textbf{Publication}
& \rotb{Architecture}
& \rotb{Set / Range}
& \rotb{Granularity}
& \rotb{Set / Range}
& \rotb{Liveliness}
& \rotb{Visibility}
& \rotb{Durability}
& \rotb{Non-Monotonicity}
& \textbf{Aggregation Model} \\ 

\cmidrule(r){1-2}
%\cmidrule(lr){2-2}
\cmidrule(lr){3-4}
\cmidrule(l){5-10}

\multicolumn{2}{>{\hsize=3.5\hsize}Y}{\textbf{System}}
& \multicolumn{2}{>{\hsize=1.5\hsize}Y}{\textbf{Feedback}}
& \multicolumn{6}{>{\hsize=6\hsize}Y}{\textbf{Reputation}} \\
\midrule

\multicolumn{10}{
>{\hsize=11\hsize}Y}{\textbf{SMPC-based Systems }} \\ 

\midrule

\ayc{Pavlov et al.}{2004}{Pavlovetal2004Supporting}
& D
& $\mathbb{R}$
& M
& ~~\,$\mathbb{R}$\newline$[0,1]$
& \faCircle
& L
& \faCircleO
& \faCircle
& Sum, Beta reputation \\ 

\mymidrule

\ayc{Yao et al.}{2007}{Yaoetal2007Private}
& D
& $\mathbb{Z}$
& M
& $\mathbb{Z}$
& \faCircle
& L
& \faCircleO
& \faCircleO
& Weighted average \\ 

\mymidrule

\ayc{Gudes et al.}{2009}{Gudesetal2009Methods}
& D
& $\mathbb{R}$
& M
& $\mathbb{R}$
& \faCircle
& L
& \faCircleO
& \faCircle
& Weighted sum, Mean \\ 

\mymidrule

\ayc{Melchor et al.}{2009}{Melchoretal2009A}
& D
& $\mathbb{Z}$
& M
& $\mathbb{Z}$
& \faCircle
& L
& \faCircleO
& \faCircleO
& Weighted average \\ 

\mymidrule

\ayc{Nithyanand and Raman}{2009}{Nithyanandetal2009Fuzzy}
& D
& ~~~$\mathbb{R}$\newline$\{0,1\}$
& M
& $\mathbb{R}$
& \faCircle
& L
& \faCircleO
& \faCircle
& Ordered weighted average \\ 

\mymidrule

\ayc{Gal-Oz et al.}{2010}{Gal-Ozetal2010Sharing}
& D
& $\mathbb{R}$
& M
& $\mathbb{R}$
& \faCircle
& L
& \faCircleO
& \faCircle
& Weighted sum, Mean \\ 

\mymidrule

\ayc{Dolev et al.}{2014}{Dolevetal2014Efficient}
& D
& $\{1,...,10\}$
& M
& $\mathbb{R}$
& \faCircle
& L
& \faCircleO
& \faCircle
& Weighted mean \\ 

\mymidrule

\ayc{Clark et al.}{2016}{Clarketal2016Dynamic}
& D
& $[0,v_{max}]$
& M
& $[0,v_{max}]$
& \faCircle
& L
& \faCircleO
& \faCircle
& Mean \\ 

\mymidrule

\ayc{Azad et al.}{2018}{Azadetal2017M2M-REP:}
& D
& $\{-1,0,1\}$
& M
& $\{-1,0,1\}$
& \faCircle
& G
& \faCircle
& \faCircle
& Weighted sum \\ 

\mymidrule

\ayc{Bakas et al.}{2020}{Bakasetal2021Functional}
& D
& $\{1,...,5\}$
& M
& $\mathbb{Z}$
& \faCircle
& G
& \faCircleO
& \faCircleO
& Weighted sum \\ 

\midrule

 \multicolumn{10}{
>{\hsize=11\hsize}Y}{\textbf{Token-based Systems}} \\ 

\midrule

\ayc{Singh and Liu}{2003}{Singhetal2003TrustMe:}
& H
& ?
& M
& ?
& ?
& G
& \faCircle
& ?
& ? \\ 

\mymidrule

\ayc{Voss}{2004}{Voss2004Privacy}
& C
& $\mathbb{Z}$
& S
& $\mathbb{Z}$
& \faCircle
& G
& \faCircle
& \faCircle
& Sum \\ 

\mymidrule

\ayc{Androulaki et al.}{2008}{Androulakietal2008Reputation}
& C
& $\{0,1\}$
& S
& $\mathbb{Z}$
& \faCircleO
& G
& \faCircle
& \faCircleO
& Sum \\ 

\mymidrule

\ayc{Kerschbaum}{2009}{Kerschbaum2009A}
& C
& $\{0,1\}$
& S
& $[0,1]$
& \faCircle
& G
& \faCircle
& \faCircle
& Beta reputation \\ 

\mymidrule

\ayc{Schiffner et al.}{2009}{Schiffneretal2009Privacy}
& C
& $\{-1,1\}$
& S
& $\mathbb{Z}$
& \faCircle
& G
& \faCircle
& \faCircle
& Sum \\ 

\mymidrule

\ayc{Hussain and Skillicorn}{2011}{Hussainetal2011Mitigating}
& C
& ?
& M
& ?
& ?
& G
& \faCircle
& ?
& Open \\ 

\mymidrule

\ayc{Schiffner et al.}{2011}{Schiffneretal2011Privacy}
& C
& $\{-,+\}$
& S
& $\mathbb{R}$
& \faCircle
& G
& \faCircle
& \faCircle
& Open \\ 

\mymidrule

\ayc{Zhang et al.}{2014}{Zhangetal2014A}
& H
& ?
& S
& $\mathbb{R}$
& \faCircle
& G
& \faCircle
& \faCircle
& Open \\ 

\mymidrule

\ayc{Bazin et al.}{2016}{Bazinetal2016A}
& H
& ?
& S
& ?
& ?
& G
& \faCircle
& ?
& Open, Beta reputation \\ 

\mymidrule

\ayc{Busom et al.}{2017}{Busometal2017A}
& C
& Text
& S
& ?
& \faCircle
& G
& \faCircle
& \faCircle
& Union \\ 

\mymidrule

\ayc{Blömer et al.}{2018}{Blomeretal2018Practical}
& C
& ?
& S
& ?
& ?
& G
& ?
& ?
& Open \\ 

\mymidrule

\ayc{Liu and Manulis}{2019}{Liuetal2019pRate:}
& C
& $\{1,...,5\}$
& M
& $\mathbb{Z}$
& \faCircle
& G
& \faCircle
& \faCircleO
& Sum \\ 

\midrule

 \multicolumn{10}{
>{\hsize=11\hsize}Y}{\textbf{Proxy-based Systems}} \\ 

\midrule

\ayc{Ries et al.}{2011}{Riesetal2011Learning}
& C
& $\{0,1\}$
& M
& $[0,1]$
& \faCircle
& L
& \faCircleO
& \faCircle
& Beta reputation \\ 

\mymidrule

\ayc{Petrlic et al.}{2014}{Petrlicetal2014Privacy-Preserving}
& C
& ~Vector\newline$\{0,1\}$
& S
& $\mathbb{Z}$
& \faCircle
& G
& \faCircle
&
& Sum \\ 

\midrule

\multicolumn{10}{
    >{\hsize=11\hsize}r}{\textit{Continues on next page}} \\

\end{tabularx}
\normalsize
\label{tab:rw}
\end{table}

%
%
% Page Break
%
%
% TABLE PART 2

\begin{table}[ht]
\tiny
\centering
\begin{tabularx}{\textwidth}{
    @{\hspace{1.1em}}>{\hangindent=1em\raggedright\hsize=3\hsize}Y
    >{\hsize=.5\hsize}Y
    >{\hsize=1\hsize}Y
    >{\hsize=.5\hsize}Y
    >{\hsize=1\hsize}Y
    >{\hsize=.5\hsize}Y
    >{\hsize=.5\hsize}Y
    >{\hsize=.5\hsize}Y
    >{\hsize=.5\hsize}Y
    >{\hsize=2.9\hsize}Y
    }

\multicolumn{10}{
    >{\hsize=11\hsize}l}{\textit{Continued from previous page}} \\ 

    \midrule
    \textbf{Publication}
    & \rotb{Architecture}
    & \rotb{Set / Range}
    & \rotb{Granularity}
    & \rotb{Set / Range}
    & \rotb{Liveliness}
    & \rotb{Visibility}
    & \rotb{Durability}
    & \rotb{Non-Monotonicity}
    & \textbf{Aggregation Model} \\ 
    
    \cmidrule(r){1-2}
    %\cmidrule(lr){2-2}
    \cmidrule(lr){3-4}
    \cmidrule(l){5-10}
    
    \multicolumn{2}{>{\hsize=3.5\hsize}Y}{\textbf{System}}
    & \multicolumn{2}{>{\hsize=1.5\hsize}Y}{\textbf{Feedback}}
    & \multicolumn{6}{>{\hsize=6\hsize}Y}{\textbf{Reputation}} \\
    \midrule

\multicolumn{10}{
    >{\hsize=11\hsize}Y}{\textbf{Signature-based Systems}} \\ 
    
    \midrule
    
    \ayc{Bethencourt et al.}{2010}{Bethencourtetal2010Signatures}
    & H
    & $\{0,1\}$
    & S
    & $\mathbb{Z}$
    & \faCircle
    & G
    & \faCircle
    & \faCircleO
    & Sum \\ 
    
    \mymidrule
    
    \ayc{Lajoie-Mazenc et al.}{2015}{Lajoie-Mazencetal2015Efficient}
    & H
    & ~$\{-,+\}$\newline$\mathbb{Z}$
    & S
    & $\mathbb{R}$
    & \faCircle
    & G
    & \faCircle
    & \faCircle
    & Open \\ 
    
    \midrule
    
     \multicolumn{10}{
    >{\hsize=11\hsize}Y}{\textbf{Transitory Pseudonym-based Systems }} \\ 
    
    \midrule
    
    \ayc{Miranda and Rodrigues}{2006}{Mirandaetal2006A}
    & C
    & ?
    & S
    & ?
    & \faCircle
    & G
    & \faCircle
    & \faCircle
    & Open \\ 
    
    \mymidrule
    
    \ayc{Steinbrecher}{2006}{Steinbrecher2006Design}
    & C
    & ?
    & S
    & ?
    & \faCircle
    & G
    & \faCircle
    & \faCircle
    & Open \\ 
    
    \mymidrule
    
    \ayc{Hao et al.}{2007}{Haoetal2007A}
    & D
    & $\{-1,1\}$
    & S
    & $\mathbb{Z}$
    & \faCircle
    & G
    & \faCircle
    & \faCircle
    & Sum \\ 
    
    \mymidrule
    
    \ayc{Hao et al.}{2008}{Haoetal2008A}
    & C
    & $\{-1,1\}$
    & M
    & $\mathbb{R}$
    & \faCircle
    & G
    & \faCircle
    & \faCircle
    & Sum, Average \\ 
    
    \mymidrule
    
    \ayc{Wei and He}{2009}{Weietal2009A}
    & C
    & $\{-1,1\}$
    & M
    & $\mathbb{R}$
    & \faCircle
    & G
    & \faCircle
    & \faCircle
    & Sum, Average \\ 
    
    \mymidrule
    
    \ayc{Peng et al.}{2010}{Pengetal2010Low}
    & C
    & ?
    & S
    & ?
    & ?
    & G
    & \faCircle
    & ?
    & Open \\ 
    
    \mymidrule
    
    \ayc{Anceaume et al.}{2013}{Anceaumeetal2014Extending}
    & D
    & $[0,1]$
    & S
    & $[0,1]$
    & \faCircle
    & G
    & ?
    & \faCircle
    & Beta reputation \\ 
    
    \mymidrule
    
    \ayc{Christin et al.}{2013}{Christinetal2013IncogniSense:}
    & C
    & ?
    & S
    & ?
    & \faCircle
    & G
    & \faCircle
    & \faCircle
    & Open \\ 
    
    \mymidrule
    
    \ayc{Clauß et al.}{2013}{Claussetal2013k-Anonymous}
    & ?
    & $[1,...)$
    & M
    & $[1,...)$
    & ?
    & G
    & ?
    & ?
    & Open \\ 
    
    \midrule
    
     \multicolumn{10}{
    >{\hsize=11\hsize}Y}{\textbf{Blockchain-based Systems }} \\ 
    
    \midrule
    
    \ayc{Schaub et al.}{2016}{Schaubetal2016A}
    & D
    & $\mathbb{Z}$
    & S
    & $\mathbb{R}$
    & \faCircle
    & G
    & \faCircleO
    & \faCircle
    & Open \\ 
    
    \mymidrule
    
    \ayc{Soska et al.}{2016}{Soskaetal2016Beaver:}
    & D
    & ~\,Text\newline$\mathbb{Z}$
    & M
    & ?
    & ?
    & G
    & \faCircle
    & ?
    & Open \\ 
    
    \mymidrule
    
    \ayc{Bazin et al.}{2017}{Bazinetal2017Self-reported}
    & D
    & $\mathbb{Z}$
    & S
    & $\mathbb{R}$
    & \faCircle
    & G
    & \faCircle
    & \faCircle
    & Open \\ 
    
    \mymidrule
    
    \ayc{Azad et al.}{2018}{Azadetal2018PrivBox:}
    & D
    & $\{-,+\}$
    & S
    & $\mathbb{Z}$
    & \faCircle
    & G
    & \faCircleO
    & \faCircle
    & Beta reputation \\ 
    
    \mymidrule
    
    \ayc{Bag et al.}{2018}{Bagetal2018A}
    & D
    & $\{0,1\}$
    & M
    & $[1,10]$
    & \faCircleO
    & L
    & \faCircleO
    & \faCircle
    & Mean \\ 
    
    \mymidrule
    
    \ayc{Bemmann et al.}{2018}{Bemmannetal2018Fully-Featured}
    & C
    & $\{0,1\}$
    & M
    & ?
    & ?
    & G
    & \faCircle
    & ?
    & ? \\ 
    
    \mymidrule
    
    \ayc{Owiyo et al.}{2018}{Owiyoetal2018Decentralized}
    & D
    & ?
    & S
    & ?
    & ?
    & G
    & \faCircle
    & ?
    & Open \\ 
    
    \mymidrule
    
    \ayc{Liu et al.}{2019}{Liuetal2019Anonymous}
    & C
    & $[1,10]$
    & S
    & $\mathbb{N}$
    & \faCircle
    & G
    & \faCircle
    & \faCircleO
    & Sum \\ 
    
    \mymidrule
    
    \ayc{Malik et al.}{2019}{Maliketal2019TrustChain:}
    & D
    & ?
    & M
    & ?
    & ?
    & G
    & \faCircle
    & ?
    & Mean, Median, Beta reputation \\ 
    
    \mymidrule
    
    \ayc{Zhou et al.}{2021}{Zhouetal2021Blockchain-based}
    & D
    & $[-5,-1]$,\newline$[1,5]$
    & M
    & $[0,1]$
    & \faCircleO
    & G
    & \faCircle
    & \faCircle
    & Weighted sum \\ 
    
    \mymidrule
    
    \ayc{Arshad et al.}{2022}{Arshadetal2022REPUTABLEA}
    & D
    & $\{0,1\}$
    & M
    & $\mathbb{R}$
    & \faCircleO
    & G
    & \faCircle
    & \faCircle
    & Beta reputation, Open \\ 
    
    \mymidrule
    
    \ayc{Putra et al.}{2022}{Putraetal2022DeTRM:}
    & D
    & $\{-1,1\}$
    & M
    & $\mathbb{R}$
    & \faCircle
    & G
    & \faCircle
    & \faCircle
    & Weighted average \\ 
    
    \midrule
    
     \multicolumn{10}{
    >{\hsize=11\hsize}Y}{\textbf{Other Systems}} \\ 
    
    \midrule
    
    \ayc{Kinateder and Pearson}{2003}{Kinatederetal2003A}
    & D
    & $[0,1]$
    & S
    & $\mathbb{R}$
    & \faCircle
    & L
    & \faCircleO
    & \faCircle
    & Open \\ 
    
    \mymidrule
    
    \ayc{Bo et al.}{2007}{Boetal2007A}
    & H
    & ?
    & S
    & ?
    & \faCircle
    & G
    & \faCircle
    & \faCircle
    & Open \\

    \mymidrule

    \ayc{Yang et al.}{2023}{Yangetal2023Towards} 
    & C 
    & $\{-1,1\}$ 
    & M 
    & $\mathbb{Z}$ 
    & \faCircle 
    & G 
    & \faCircle 
    & \faCircle 
    & Sum \\
    
    \bottomrule    

\end{tabularx}
\end{table}

\end{document}